# Synthetic magnetoelectric coupling in a nanocomposite multiferroic

P. Jain,[1] Q. Wang,[1] M. Roldan,[2] A. Glavic,[3] V. Lauter,[3] C. Urban,[4] Z. Bi,[1] T. Ahmed,[1] J. Zhu,[1] M. Varela,[2,3] Q. Jia,[1] M.R. Fitzsimmons*[1]

[1]Los Alamos National Laboratory, Los Alamos NM 87545; [2]Universidad Complutense de Madrid, Madrid Spain 28040; [3]Oak Ridge National Laboratory, Oak Ridge TN 37831; [4]University of California at San Diego, La Jolla CA 92093
*Correspondence and requests for materials should be addressed to M.R.F. (fitz@lanl.gov)

**Given the paucity of single phase multiferroic materials (with large ferromagnetic moment), composite systems seem an attractive solution in the quest to realize magnetoelectric coupling between ferromagnetic and ferroelectric order parameters. Despite having antiferromagnetic order, $BiFeO_3$ (BFO) has nevertheless been a key material in this quest due to excellent ferroelectric properties at room temperature. We studied a superlattice composed of 8 repetitions of 6 unit cells of $La_{0.7}Sr_{0.3}MnO_3$ (LSMO) grown on 5 unit cells of BFO. Significant net uncompensated magnetization in BFO is demonstrated using polarized neutron reflectometry in an insulating superlattice. Remarkably, the magnetization enables magnetic field to change the dielectric properties of the superlattice, which we cite as an example of synthetic magnetoelectric coupling. Importantly, this controlled creation of magnetic moment in BFO suggests a much needed path forward for the design and implementation of integrated oxide devices for next generation magnetoelectric data storage platforms.**

The ability to control magnetization, M, via electric fields or alternatively electric polarization, P, via magnetic fields enables a myriad of technological innovations in information storage, sensing, and computing. For example, Oersted-fields that are presently used to switch the magnetic state of commercial magnetic tunnel junctions, are spatially extended and require modest current to produce. These attributes limit the areal density of magnetic tunnel junctions. Because electrostatic fields can be confined and require very little current to produce, integration of a multiferroic composite—a system of different constituents with coupled M and P order parameters—into a magnetic tunnel junction might enable the "single memory solution"—non-volatile memory that is more energy efficient, faster, higher capacity and more affordable than competing technologies.

$BiFeO_3$ (BFO) is a single phase multiferroic material which exhibits magnetoelectric coupling between antiferromagnetic[1] and ferroelectric[2] order parameters to temperatures hundreds of degrees above room temperature. As such, BFO is potentially an attractive technological material. However, important challenges impede progress. First, the electric polarization vector can be along any of eight equivalent [111] directions, thus, the polarization domain state can be ill-defined/complex.[3] Second, the sub-lattice magnetization has six equivalent easy axes in the plane normal to the electric polarization vector, thus, the antiferromagnetic domain state is ill-defined even if the polarization were saturated.[3] Third, because BFO is an antiferromagnet, there is virtually no net moment[4,5] which can interact with an applied magnetic field, *i.e.,* BFO lacks a "magnetic handle". Calculations performed by Ederer and Spaldin[6] suggest that canting of the antiferromagnetic structure due to the Dzyaloshinskii-Moriya (DM) interaction could produce a small uncompensated moment of $0.1\mu_B$/Fe (~15kA/m) in BFO—a magnetization about two orders smaller than what is found in a magnetic tunnel junction. The DM interaction is thus unlikely to produce a magnetic handle.

Various groups[7,8,9,10,11] have explored attaching a magnetic handle to an AFM (e.g., BFO) using exchange bias.[12] Exchange bias is the shift of the ferromagnetic hysteresis loop about zero applied magnetic field that can be observed for ferromagnetic/antiferromagnetic composites. In the case of a ferromagnet (FM) deposited on BFO, electric fields could change the magnetization of the FM via a change of AFM structure—the latter is magnetoelectrically coupled to electric polarization in BFO. The FM/BFO layered-composite would thus exhibit a large net magnetization and (synthetic) magnetoelectric multiferroic coupling though not in the same sense as a single phase material.

The approach has met with some success; however, complete switching of the *saturation* magnetization with an electric field has proven elusive. To date the magnitude of exchange bias, $|H_E|$ has been smaller than the coercive field, $H_c$, of the FM, so the saturation magnetization cannot be fully reversed at zero magnetic field. Because $H_E$ is proportional to the component of the sub-lattice magnetization parallel to the FM magnetization,[12] the ill-defined antiferromagnetic domain state of BFO, even when the electric polarization state is well-defined, invariably compromises $|H_E|$.

Here, we propose a much different approach to realize a magnetoelectric multiferroic. Namely, we demonstrate intimate coupling (though not exchange bias) between the magnetization of LSMO with the *uncompensated* magnetization of BFO layers in a superlattice structure. The size of the uncompensated magnetization suggests that when a very thin BFO film is sandwiched between LSMO, BFO becomes *ferri*-magnetic. The temperature dependence of the ferrimagnetic order parameter in BFO is the same as that of the ferromagnetic order parameter of LSMO, suggesting that LSMO induces the uncompensated magnetization in BFO. In addition, we demonstrate control over the dielectric constant of the superlattice with *magnetic* field. We also propose a means to extend our discovery above room temperature.

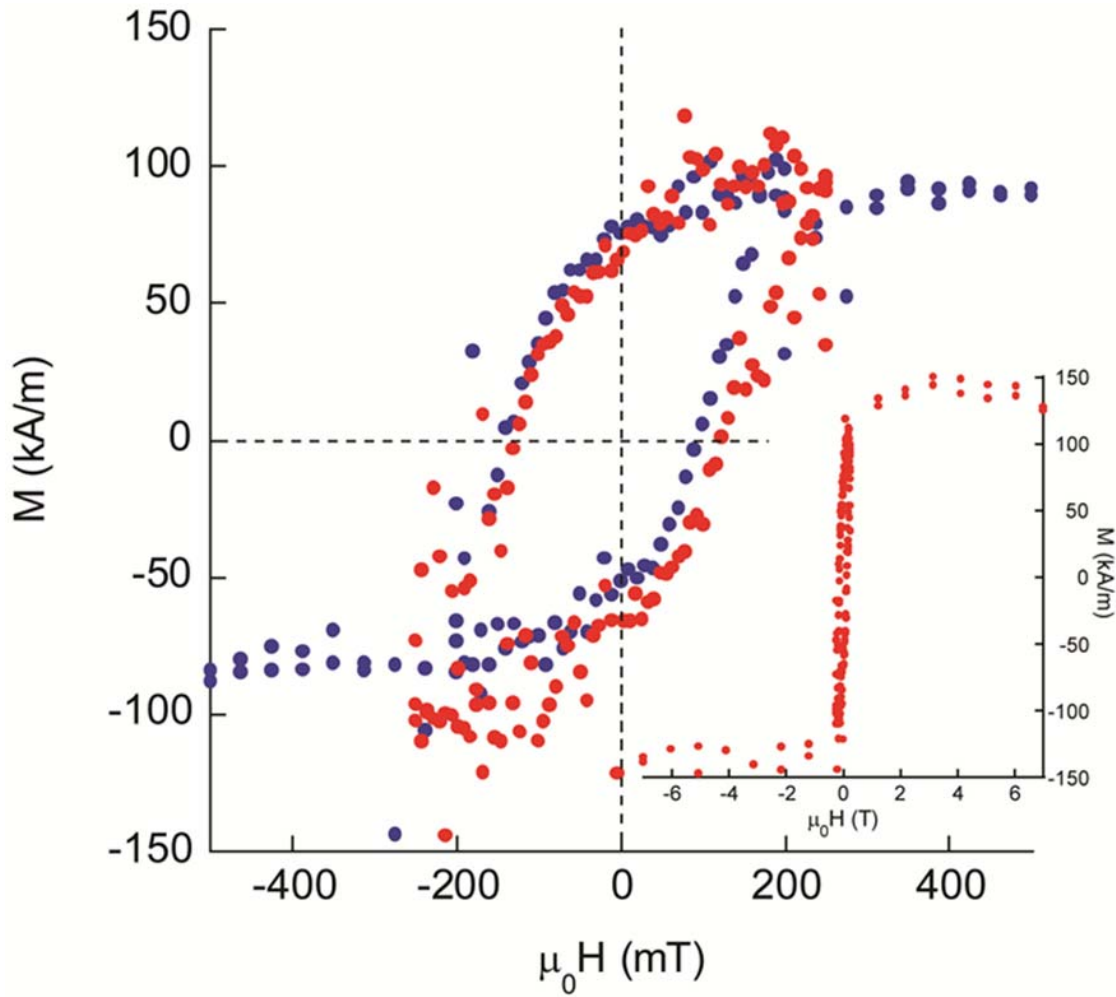

**Figure 1** Magnetization vs. magnetic field for superlattice measured from (blue) ±0.5 T and (red, and inset) ± 7 T at 10 K. This measurement demonstrates the loop shift in blue is an artifact of a minor loop of the hysteresis (in red).

We grew a $[(LSMO)_n/(BFO)_m]_N$ superlattice on a (001) $SrTiO_3$ (STO) substrate by pulsed laser (KrF) deposition, where n = 6, m = 5 unit cells and N = 8 is then number LSMO/BFO bilayers. Evidence for chemically and structurally well-defined interfaces over lateral dimensions of tens of nm was obtained using high angle annular dark field (HAADF) Z-contrast microscopy [Fig. S1 of Ref. 13] and x-ray reflectometry [Fig. S2 of Ref. 13]. The LSMO and BFO layers are single crystals and microscopy of the BFO/STO interface indicate excellent epitaxy.[14] Electron energy loss spectroscopy across the entire thickness over a 5 nm wide region of the superlattice sample found no evidence for $Fe^{2+}$. Further, we found no evidence for structural phases other than LSMO, BFO and STO using x-ray diffraction [Fig. S3 of Ref. 13]. Two bilayers are somewhat rougher than the other six bilayers. Because our motivation for growing a superlattice was to augment the signal from neutron scattering, we expect the neutron scattering (as well as the transport measurements) to be representative of the majority of the sample. We also grew a 20 nm thick single crystalline BFO film on STO. Later we discuss comparisons of neutron and capacitance data taken from the superlattice and BFO film samples.

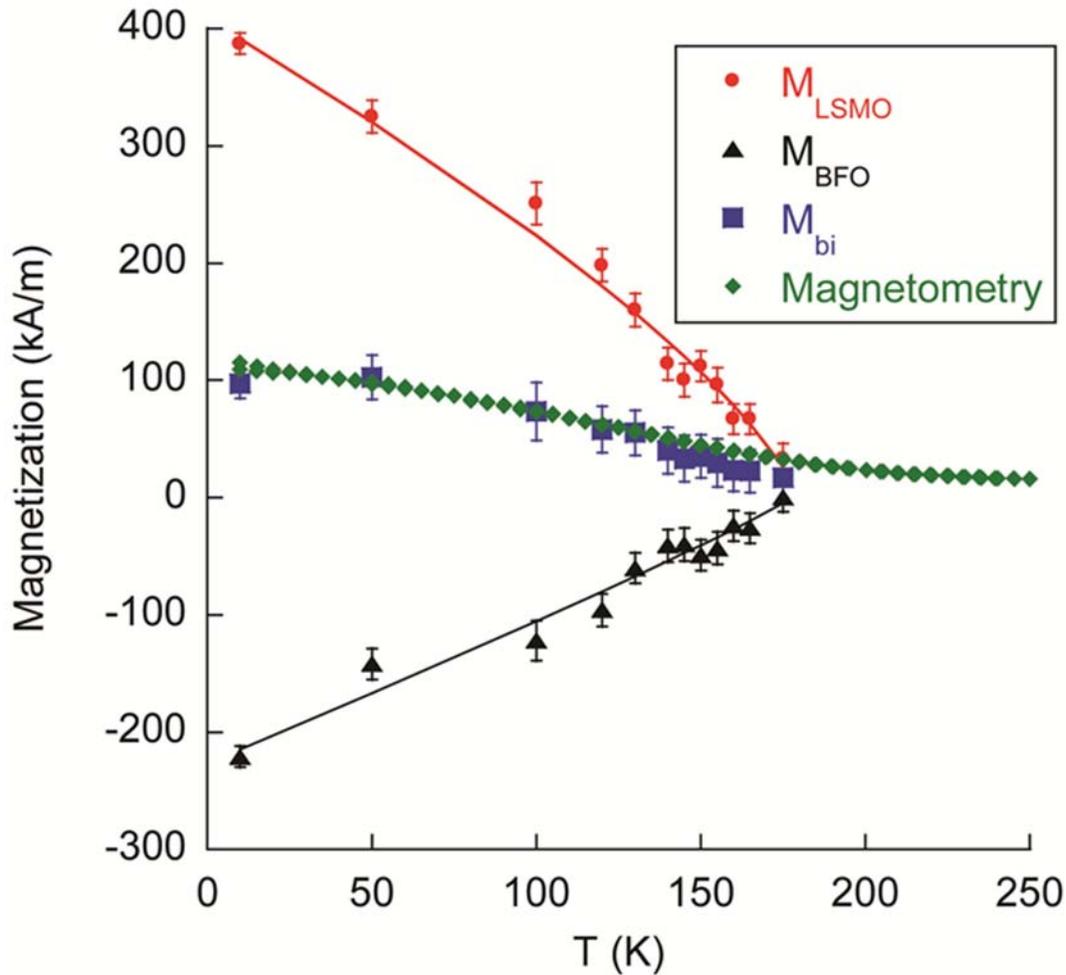

**Figure 2** The temperature dependence of the magnetizations for LSMO (circles) and BFO (triangles) layers in the superlattice, (squares) the thickness-weighted average of these magnetization and (diamonds) the moment of the sample measured with magnetometry normalized by the volume of the superlattice film.

We measured the magnetic and electronic properties of the superlattice sample after cooling in a magnetic field of 0.5 T to 10 K. The field was applied along [100] of STO. The blue symbols in Fig. 1 show the magnetization of the sample, M, versus applied field, $\mu_0 H$, after cycling ± 0.5 T. The hysteresis loop is shifted by -20 mT. The shift is three times larger than that observed by Wu et al.[11] In order to determine whether the origin of the loop shift was due to exchange bias (a consequence of unidirectional anisotropy) or a minor loop (failure to completely saturate the magnetization)[15,16] we repeated the measurement cycling $\mu_0 H$ from ±7 T. The result is shown by the red symbols (Fig. 1 and inset). The loop shift of the red colored loop (-2 mT) is not significantly different from zero. Thus, the superlattice does not exhibit exchange bias, although the LSMO and BFO layers can still be exchange coupled (but not in a manner that produces unidirectional anisotropy). The absence of exchange bias is expected since the BFO thickness in the superlattice is much less than the critical thickness >10 nm required to establish unidirectional anisotropy in BFO.[9]

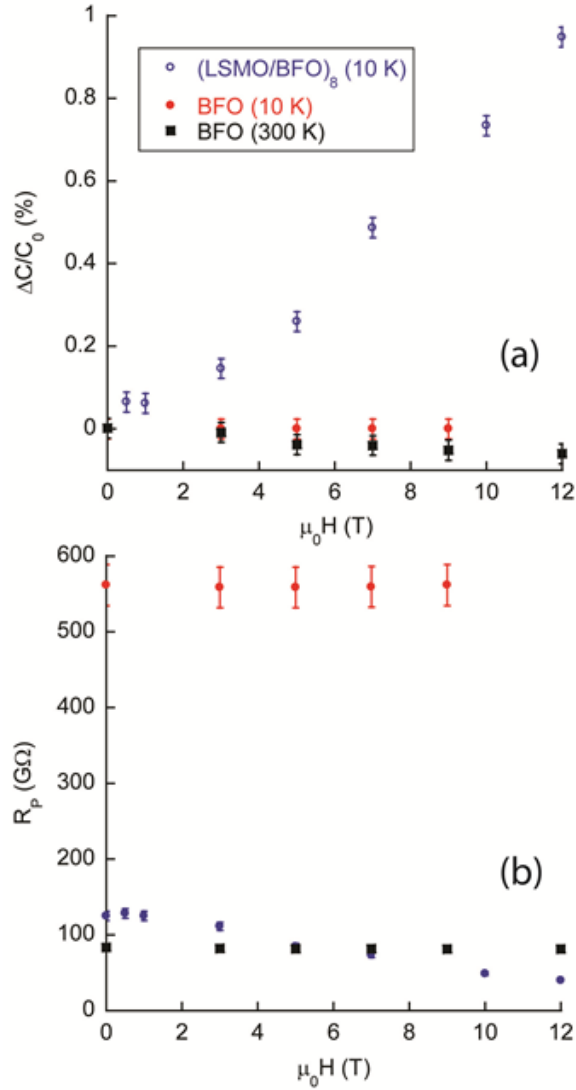

**Figure 3(a) Change of capacitance normalized to the capacitance at zero magnetic field vs. magnetic field for the superlattice and 20 nm thick BFO film. (b) In-plane resistance of the superlattice and the 20 nm thick BFO film vs. magnetic field. For 10 K and $\mu_0 H$ = 0 T, the resistivity of the superlattice sample is (7.5±0.3)x$10^5$ Ωcm and (1.1±0.1)x$10^6$ Ωcm for the thick BFO film.**

Polarized neutron reflectivity data of the superlattice were acquired at the Spallation Neutron Source, Oak Ridge National Laboratory. Briefly, the reflectivity, R, was measured with two neutron beam polarizations—one parallel (+) and one opposite (-) to the 0.5 T applied field as a function of wavevector transfer, Q, (the difference between the incident and specularly reflected neutron wavevectors). $R^{\pm}(Q)$ (Fig. S4 of Ref. 13) was measured to 0.2 Å$^{-1}$ at 10 K after field cooling the sample described previously. Guided by results from x-ray reflectometry, a model of the chemical and magnetic structure of the superlattice was fitted to the data. From this analysis the magnetizations representative of the LSMO and BFO layers were obtained. The temperature dependencies of the LSMO (circles) and BFO (triangles) magnetizations are shown in Fig.

2. There are three essential observations. (1) The magnetization of BFO (equivalent to ~1.3 $\mu_B$/Fe at 10 K) is much larger than can be attributed to the DM interaction,[6] although similarly large moments have been found in BFO films previously.[17] (2) The uncompensated BFO magnetization is opposite to the LSMO magnetization and the applied field. [Note, the agreement between the thickness weighted magnetization of LSMO and BFO from neutron scattering (squares, Fig. 2) agree very well with the magnetization measured using magnetometry (diamonds, Fig. 2), thus, further confirming the anti-parallel orientation of LSMO and BFO magnetizations.] (3) The thermal dependence of the BFO and LSMO magnetic order parameters are the same.

Whereas Fig. 2 shows three remarkable observations about the magnetic structure of the superlattice, Fig. 3(a) shows an equally remarkable feature—the influence of magnetic field on the capacitance of the LSMO/BFO superlattice. Capacitance was measured using an Andeen-Hagerling Capacitance Bridge with potential applied along the sample plane. At 10 K, the capacitance of the superlattice increases with $\mu_0 H$ at a rate of 0.1%/T. In contrast, the capacitance of the 20 nm thick BFO film is unchanged with field at 10 and 300 K. Both samples were insulating during the measurements [Fig. 3(b)]. Our conclusion from the data in Fig. 3 is that we can control the capacitance of a LSMO/BFO superlattice using *magnetic* field.

Polarized neutron reflectivity measurements of the 20 nm thick BFO film (measurements at the Los Alamos Neutron Science Center) detected no neutron spin dependence of $R^{\pm}(Q)$ (Fig. S5 of Ref. 13). Thus, the large uncompensated magnetization of BFO in the superlattice is a consequence of unique features associated with the superlattice, *e.g.,* its growth, strain, architecture, proximity to a ferromagnet, *etc.*

The origin of uncompensated magnetization in BFO films has been the subject of some controversy.[17,18,19] One explanation attributes the uncompensated magnetization in BFO thin films to epitaxial strain.[17,19] Another explanation attributes the magnetization to oxygen deficiency leading to a change of Fe valence from +3 to +2.[18] $Fe^{2+}$ tends to make the film metallic, thus, compromising magnetoelectric coupling[18] and tempering interest in BFO as a technological material. More recently, Borisevich *et al.*[20] have observed that oxygen octahedral tilts, normally present in BFO, are suppressed within the first 3 to 4 unit cells of BFO in proximity to LSMO. The spin structure of BFO in which oxygen octahedral tilts are suppressed is unknown. Because the thickness of BFO layers in our superlattice is 5 unit cells, every BFO unit cell is within 3-4 unit cells of LSMO. We expect oxygen octahedral tilts are likely suppressed in our superlattice. Accordingly, we suggest regardless of the presence of $Fe^{2+}$, epitaxial strain or suppression of tilts, the electronic and magnetic structures and properties of BFO in the superlattice are likely different than those of the 20 nm thick BFO film or bulk BFO. We have explored computational models of BFO/LSMO heterostructures with thinner layers of BFO and LSMO (1-3 unit cells) and tilt-free octahedra in BFO along the growth direction. Uncompensated magnetization on the BFO-side of the BFO/LSMO interface was found using density functional theory.[21]

Termination of the LSMO layer, $MnO_2$ vs. SrO, influences the sign of exchange coupling across the LSMO/BFO interface.[22] Specifically, spins of $MnO_2$ terminated LSMO layers are *anti*-ferromagnetically coupled to spins in the adjacent BFO layer. Our results are consistent with

*anti*-ferromagnetically coupled spins across MnO$_2$ terminated layers. Growth of LSMO films on TiO$_2$-terminated (001) STO substrates yields MnO$_2$ terminated LSMO layers;[22] however, in our superlattice LSMO layers are grown on BFO layers and the termination of these layers is unknown.

The temperature dependencies of the LSMO and BFO magnetic order parameters were fitted to the usual form $\left(1 - \frac{T}{T_c}\right)^\beta$ for a second order phase transition (curves, Fig. 2). The values of T$_c$ = 179 ± 1 K for the two films are not significantly different. Suppression of T$_c$ well below 350 K is typical for LSMO films less than 30 nm thickness.[23] The important observation is that in the absence of LSMO or magnetically ordered LSMO, we find no evidence for uncompensated magnetization in a BFO film, yet in the presence of ferromagnetic LSMO, uncompensated magnetization is induced in 5 unit cell thick BFO layers, and the temperature dependence of the magnetization is the same (and it was not constrained to be the same in fitting the neutron data) as that of the LSMO film. Thus, an intimate connection between the magnetic order parameters of LSMO and BFO exists for our superlattice. (The intimate connection between the order parameters is essential to realizing synthetic magnetoelectric coupling at room temperature which may be accomplished by increasing the thickness of *only* the LSMO layers to achieve T$_c$ ~ 350 K—typical of bulk LSMO.)

Uncompensated magnetization in the BFO component of the superlattice provides an opportunity for magnetic field to affect the electronic properties of BFO, provided the uncompensated magnetization is coupled to the sub-lattice magnetization of BFO. The observed change of capacitance with field may be due to correlation between measurement of capacitance and field-induced-change of resistance, magnetostriction, or ideally magnetoelectric coupling.

In order to determine the influence of a change of resistance on capacitance, we measured the capacitance of a test capacitor as a function of a variable resistance placed in parallel with the capacitor. We found a threefold increase in resistance produced a 0.3 % increase in the measured capacitance of the circuit. The increase is only one-fourth the variation of capacitance shown in Fig. 3a over a 12 T range of field in which the resistance of the sample also changed by a factor of three (Fig. 3b). Therefore, a change in the parallel resistance of the sample cannot account for the change of the sample's capacitance with field.

For strongly magnetic materials the magnetostriction at saturation is of order 10$^{-5}$ (i.e., the change in dimension due to alignment of domains is about 10 ppm). The saturation magnetization of the superlattice is achieved for fields of not more than 1 T, thus, magnetostrictive effects are at least two orders of magnitude too small to explain the change of capacitance (induced by the change in distance between electrodes).

We propose that magnetoelectric coupling intrinsic to BFO films and bulk BFO persists in *ferri*-magnetic ultra-thin BFO layers. An applied magnetic field aligns the net *ferri*-magnetic moment, which in turn establishes a preference for a subset of easy planes of the sub-lattice magnetization. Domains with electric polarization normal to the preferred easy planes of the sub-lattice magnetization grow at the expense of all other polarization domains. Thus, application of field alters the net electric polarization of the superlattice, which manifests itself as an increased dielectric constant in the capacitance value we measured.

In conclusion, we have developed a composite of two materials, neither independently exhibit dielectric response to magnetic field, but when fashioned into a superlattice, the dielectric constant changes by 0.1% / Tesla. The superlattice consists of 8 repetitions of 6 unit cells of LSMO grown on 5 unit cells of BFO. Below 179 K, LSMO is ferromagnetic and BFO exhibits net uncompensated magnetization with the magnetization of BFO opposite to that of the LSMO. The magnetic order parameters have the same dependence with temperature, suggesting an intimate relationship. While uncompensated magnetization in the BFO layers is intimately linked to the LSMO magnetization, its detailed origin is unknown. Nevertheless, we have discovered a new means to produce synthetic magnetoelectric coupling in a nanocomposite at 10 K. Synthetic magnetoelectric coupling might be possible at room temperature in a superlattice consisting of ~75 unit cell thick LSMO layers (to achieve $T_c$ ~ 350 K comparable to bulk LSMO) while confining BFO layers to a thickness of 5 unit cells.

## Acknowledgement

This work was supported by the LANL/LDRD program and the Center for Integrated Nanotechnologies (CINT) at Los Alamos National Laboratory. This work has benefited from the use of the Spallation Neutron Source (Oak Ridge National Laboratory) and the Lujan Neutron Scattering Center (Los Alamos National Laboratory), which are funded by the Scientific User Facilities Division of the Department of Energy's Office of Basic Energy Science. Part of the work was carried out at the National High Magnetic Field Laboratory's High B/T Facility supported by the NSF-DMR-1157490. Los Alamos National Laboratory is operated by Los Alamos National Security LLC under DOE Contract DE-AC52-06NA25396. The research at UCSD was supported by the Office of Basic Energy Science, U.S. Department of Energy, BES-DMS funded by the Department of Energy's Office of Basic Energy Science, DMR under grant DE FG03 87ER-45332. P.J. would also like to acknowledge Institute for Complex Adaptive Matter for financial support.

## Author contributions

P.J. performed the capacitance measurements, Q.W., A.G., V.L. and M.F. performed the neutron measurements and analysis, M.R. and M.V. performed electron microscopy, C.U. performed magnetometry, Z.B. and Q.J. made the samples, T.A. and J.Z. contributed theoretical insights.

**Competing financial interests**: The authors declare no competing financial interests

## References

[1] S.V. Kiselev, R.P. Ozerov and G.S. Zhdanov, Detection of magnetic order in ferroelectric $BiFeO_3$ by neutron diffraction, *Sov. Phys. Dokl.* **7**, 742 (1963).

[2] J.R. Teague, R. Gerson and W.J. James, Dielectric hysteresis in single crystal $BiFeO_3$, *Solid State Commun.* **8**, 1073 (1970).

[3] S. Lee, W. Ratcliff II, S-W. Cheong, and V. Kiryukhin, Electric field control of the magnetic state in BiFeO$_3$ single crystals, *Appl. Phys. Lett.* **92**, 192906 (2008).

[4] H. Béa *et al.,* Influence of parasitic phases on the properties of BiFeO$_3$ epitaxial thin films, *Appl. Phys. Lett.* **87**, 072508 (2005).

[5] H. Béa *et al.,* Investigation on the origin of the magnetic moment of BiFeO3 thin films by advanced x-ray characterizations, *Phys. Rev. B* **74,** 020101(R) (2006).

[6] C. Ederer and N.A. Spaldin, Weak ferromagnetism and magnetoelectric coupling in bismuth ferrite, *Phys. Rev. B* **71,** 060401 (2005).

[7] X. He, *et al.,* Robust isothermal electric control of exchange bias at room temperature, *Nature Materials,* **9**, 579 (2010).

[8] Y-H Chu, L.W. Martin, M.B. Holcomb and R. Ramesh, Controlling magnetism with multiferroics, *Materials Today,* **10**, 16 (2007).

[9] H. Béa, *et al.,* Mechanisms of exchange bias with multiferroic BiFeO$_3$ epitaxial, *Phys. Rev. Lett.,* **100,** 017204 (2008).

[10] S.M. Wu *et al.,* Reversible electric control of exchange bias in a multiferroic field-effect device, *Nature Materials,* **9**, 756 (2010).

[11] S.M. Wu *et al.,* Full electric control of exchange bias, *Phys. Rev. Lett.* **110**, 067202 (2013).

[12] J. Nogués and Ivan K. Schuller, Exchange bias, *J. of Magn. and Magn. Mater.* **192**, 203 (1999).

[13] Reference to supplementary materials.

[14] S. Singh *et al.,* Induced magnetization in La$_{0.7}$Sr$_{0.3}$MnO$_3$/BiFeO$_3$ superlattices, *Phys. Rev. Lett.* **113**, 047204 (2014).

[15] J. Nogués *et al.,* Exchange bias in nanostructures, *Physics Reports* **422**, 65 (2005).

[16] I.S. Jacobs and C.P. Bean, In Magnetism (edited by G.T. Rado, H. Suhl, Academic Press, New York), 271 (1963).

[17] J. Wang *et al.,* Epitaxial BiFeO$_3$ multiferroic thin film heterostructures, *Science* **299**, 1719 (2003).

[18] W. Eerenstein et al., Comment on "Epitaxial BiFeO$_3$ multiferroic thin film heterostructures", *Science* **307**, 1203 (2005).

[19] J. Wang, *et al.,* Response to comment on "Epitaxial BiFeO$_3$ multiferroic thin film heterostructures", *Science* **307** 1203 (2005).

[20] A.Y. Borisevich *et al.,* Suppression of octahedral tilts and associated changes in electronic properties at epitaxial oxide heterostructure interfaces, *Phys. Rev. Lett.* **105**, 087204 (2010).

[21] Towfiq Ahmed *et al.,* to be submitted (2014).

[22] P. Yu, et al., Interface control of bulk ferroelectric polarization, *Proc. Natl. Acad. of Sci.* DOI/10.1073/pnas.1117990109

[23] P. Perna *et al.,* High Curie temperature for La$_{0.7}$Sr$_{0.3}$MnO$_3$ thin films deposited on CeO2/YSZ-based buffered silicon substrates, *J. Phys.:Condens. Matter* **21**, 306005 (2009).